\documentstyle[twoside,fleqn,espcrc2]{article}
\newcommand{\half}{{\scriptstyle{{1\over 2}}}}
\newcommand{\quarter}{{\scriptstyle{{1\over 4}}}}
\def\beq{\begin{equation}}
\def\eeq{\end{equation}}
\def\bea{\begin{array}}
\def\eea{\end{array}}
\def\beqa{\begin{eqnarray}}
\def\eeqa{\end{eqnarray}}

\def\sd{self-du\-al}
\def\myre{{\rm Re}}
\def\u1{{U(1)}}
\def\su2{{SU(2)}}
\newcommand{\re}{\relax{\rm I\kern-.18em R}}
\def\cT{{\beta}}
\def\cP{{\cal{P}}}
\def\tr{{\rm tr}} 
\def\pl{{{\cal P}_\infty}}
\def\plo{{{\cal P}_\infty^0}}

\newcommand{\AmS}{{\protect\the\textfont2
  A\kern-.1667em\lower.5ex\hbox{M}\kern-.125emS}}

\title{Constituent monopoles without gauge fixing
\vskip-3cm\hfill\small INLO-PUB-10/98\vskip2.5cm
}
\author{Thomas C. Kraan\address{Instituut-Lorentz for Theoretical Physics, 
University of Leiden,\\~PO Box 9506, NL-2300 RA Leiden, The Netherlands.}
and Pierre van Baal${}^{\rm a}{}$\thanks{Presented by second author
~at Lattice '98, 13-18 July.}}
\begin{document}
\begin{abstract}
We discuss the recent construction of new exact finite temperature instanton 
solutions with a non-trivial value of the Polyakov loop at infinity. They can 
be shown, in a precise and gauge invariant way, to be formed by the 
superposition of $n$ BPS monopoles for an $SU(n)$ gauge group. 
\end{abstract}
\maketitle
\section{Introduction}
Instantons at finite temperature (or cal\-orons) are constructed on $\re^3 
\times S^1$, taking a periodic array of instantons. For $SU(2)$ the five 
parameter Harrington-Shepard solution~\cite{HarShe} can be formulated within 
the 't~Hooft ansatz. New exact solutions with a non-trivial value of
the Polyakov loop at infinity~\cite{GroPisYaf} were only constructed very 
recently, either using~\cite{LeeLu} results due to Nahm~\cite{Nahm} or by 
using~\cite{KrvB} the well-known ADHM construction~\cite{ADHM}, translated 
by Fourier transformation to the Nahm language. Thus mapped to an Abelian 
problem on the circle, the quadratic ADHM constraint is solved~\cite{KrvB}.

\section{New caloron solutions}
In the periodic gauge, $A_\mu(x\!+\!\cT)\!=\!\!A_\mu(x)$, 
the Polyakov loop at spatial infinity
\beq
\pl=\lim_{|\vec x|\rightarrow\infty}
P\,\exp(\int_0^\cT A_0(\vec x,t)dt),
\eeq
after a constant gauge transformation, is characterised by 
($\sum_{m=1}^n\mu_m=0$)
\beqa
\plo=\exp[2\pi i{\rm diag}(\mu_1,\ldots,\mu_n)],\qquad\quad&&\\
\mu_1<\ldots<\mu_n<\mu_{n\!+\!1}\equiv\mu_1+1.&&\nonumber
\eeqa

A non-trivial value, $\cP\not\in Z_n$, acts like a Higgs field. We 
found~[5c] a remarkably simple formula for the action density, valid 
for arbitrary $SU(n)$. Using the classical scale invariance to put $\beta=1$, 
\beqa
\tr F_{\mu\nu}^{\,2}=\partial_\mu^2\partial_\nu^2\log\psi,\ 
\psi\!=\!-\cos(\!2\pi t)\!+\half\tr\!\prod_{m=\!1}^n\!A_m,\hskip-5mm\nonumber
\eeqa
\beq
A_m\equiv\frac{1}{r_m}\left(\!\!\!\bea{cc}r_m\!\!&|\vec y_m\!\!-
\!\vec y_{m+1}|\\0\!\!&r_{m+1}\eea\!\!\!\right)\left(\!\!\!\bea{cc}c_m\!\!&
s_m\\s_m\!\!&c_m\eea\!\!\!\right),
\eeq
with $r_m=|\vec x-\vec y_m|$ the center of mass radius of the 
$m^{\rm th}$ constituent monopole, which can be assigned a mass 
$16\pi^2\nu_m$, where $\nu_m\equiv\mu_{m+1}-\mu_m$. Also
$r_{n+1}\equiv r_1$, 
$\vec y_{n+1}\equiv\vec y_1$, $c_m\equiv\cosh(2\pi\nu_m r_m)$ and
$s_m\equiv\sinh(2\pi\nu_m r_m)$. The order of matrix multiplication 
is crucial here, $\prod_{m=1}^n A_m\equiv A_n\ldots A_1$.
\begin{figure}[htb]
\vspace{6cm}
\includegraphics{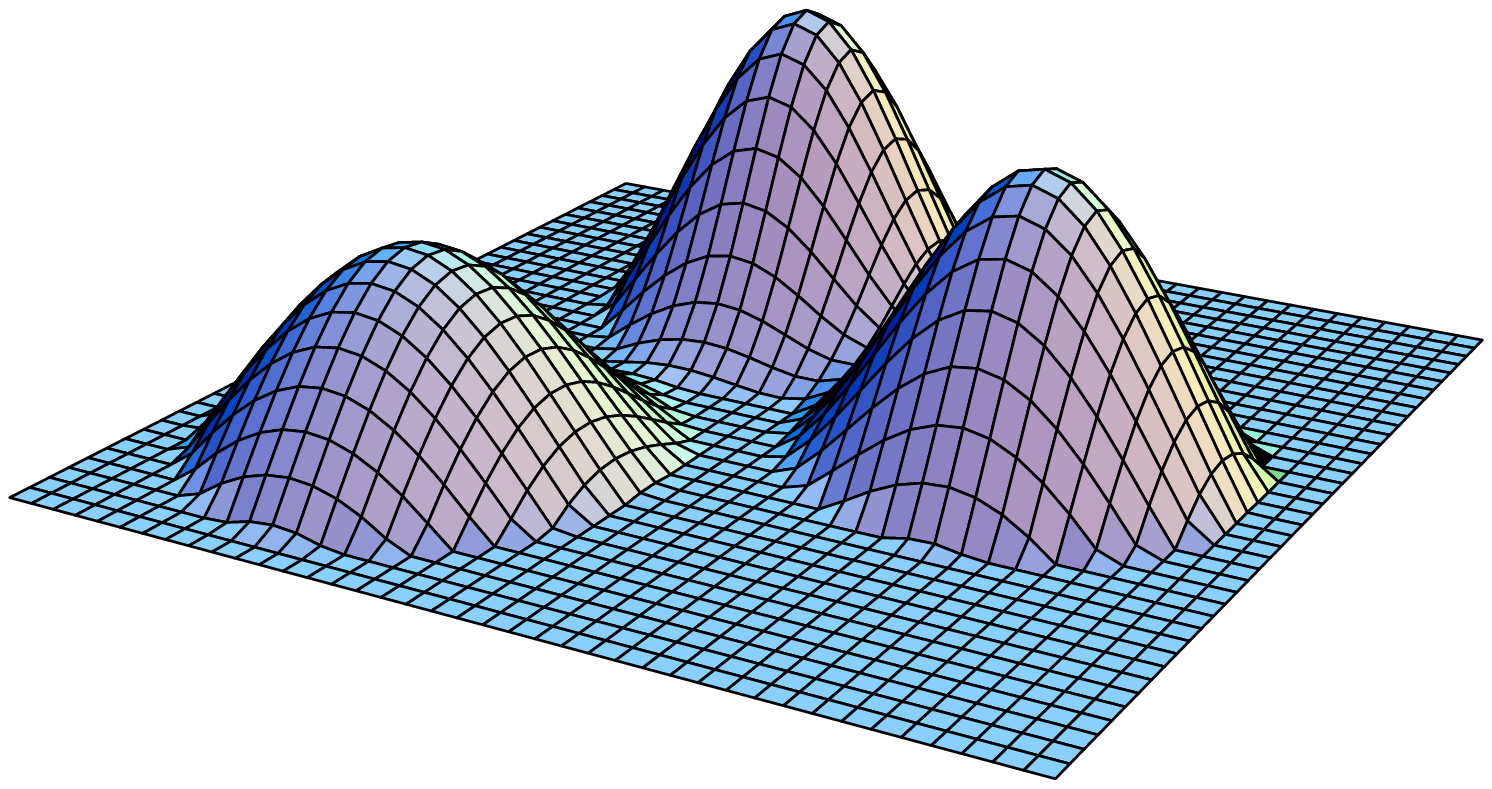}
\includegraphics{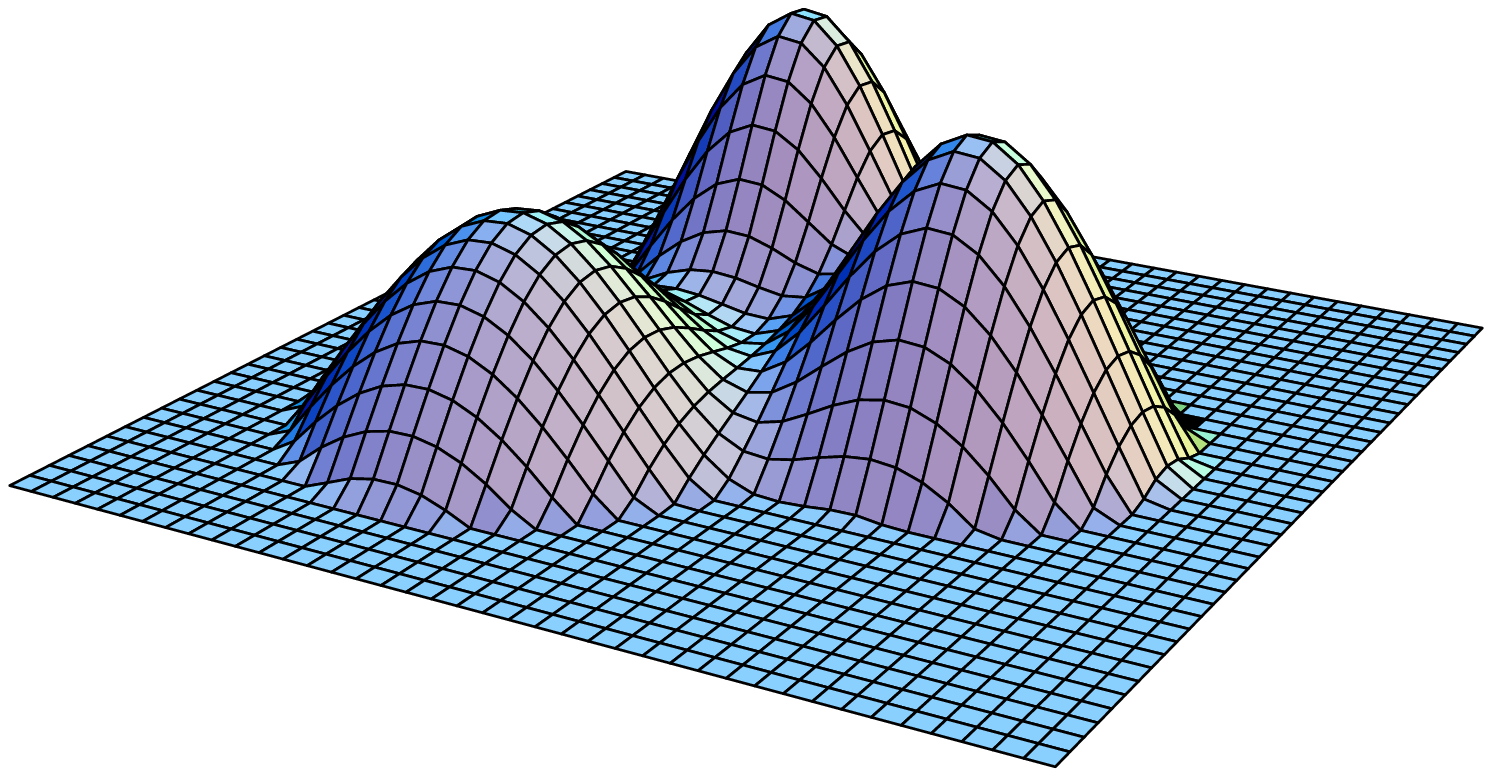}
\includegraphics{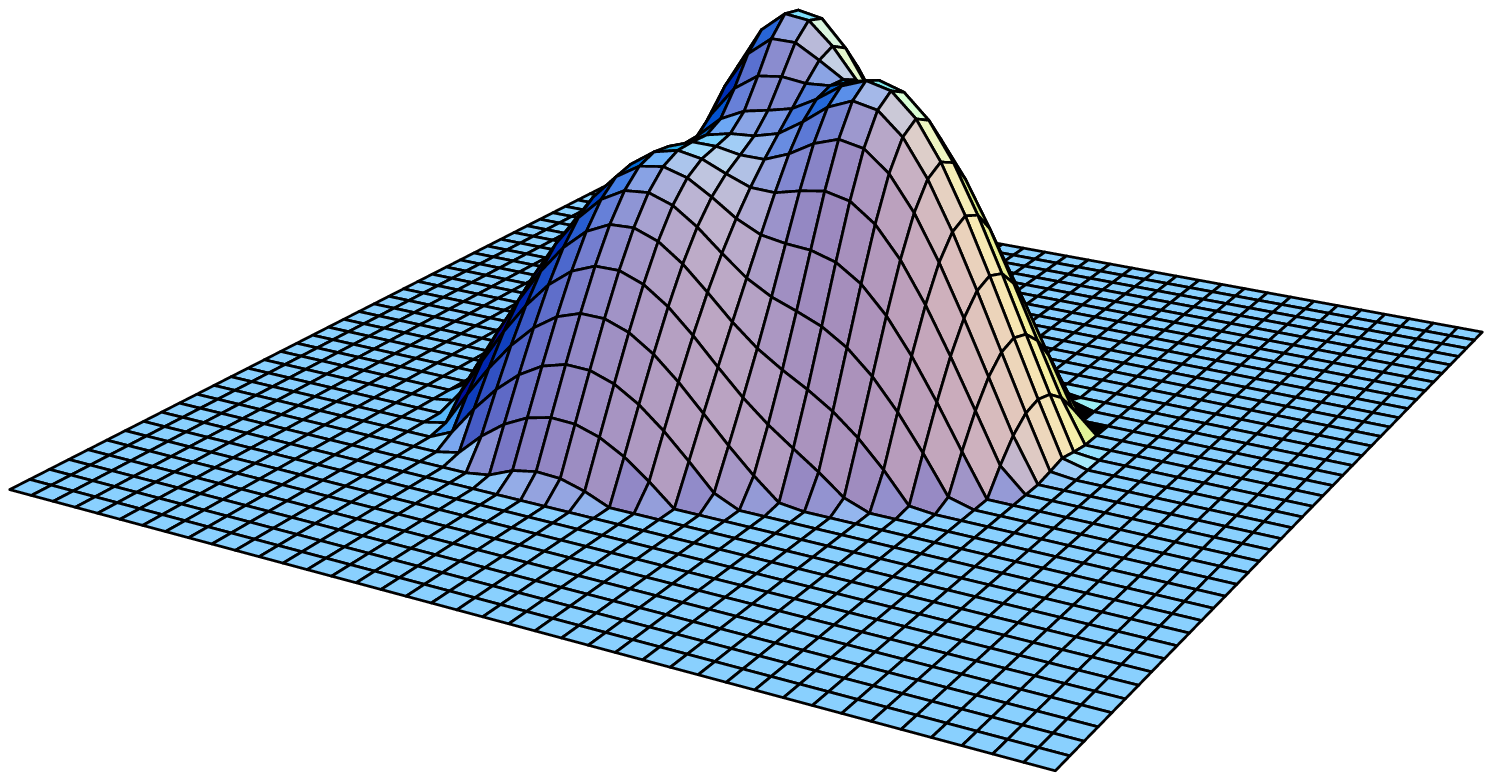}
\caption{Action densities for the $SU(3)$ caloron on equal logarithmic scales, 
cut off at $1/e$, for $t=0$ in the plane defined by $\vec y_1\!=\!(-\half,
\half,0)$, $\vec y_2=(0,\half,0)$ and $\vec y_3=(\half, -\quarter,0)$, 
in units of $\cT$, for $\cT=1/4$, $1/3$ and $2/3$ from top to bottom, 
using $(\mu_1,\mu_2,\mu_3)=(-17,-2,19)/60$.}
\end{figure}

For $\pl=\exp(2\pi i\omega\tau_3)$ the $SU(2)$ gauge field reads~[5a], in 
terms of 
the anti-selfdu\-al 't~Hooft tensor $\bar\eta$ and Pauli matrices $\tau_a$,
\beqa
&&\hskip-6mm A_\mu(x)=\frac{i}{2}\bar\eta^3_{\mu\nu}\tau_3\partial_\nu\log
\phi(x)+\\
&&\frac{i}{2}\phi(x)\myre\left((\bar\eta^1_{\mu\nu}-i\bar\eta^2_{\mu\nu})
(\tau_1+i\tau_2)\partial_\nu\chi(x)\right),\nonumber
\eeqa
where $\phi^{-1}=1-\frac{\pi\rho^2}{\psi}\left(\frac{s_1c_2}{r_1}+\frac{s_2
c_1}{r_2}+\frac{\pi\rho^2s_1s_2}{r_1r_2}\right)$ and $\chi=\frac{\pi
\rho^2}{\psi}\left(e^{-2\pi it}\frac{s_1}{r_1}+\frac{s_2}{r_2}\right)
e^{2\pi i\nu_1 t}$, with $\nu_1=2\omega$, $\nu_2=1-2\omega$ and $\pi\rho^2=
|\vec y_2-\vec y_1|$. The solution is presented in the ``algebraic'' gauge, 
$A_\mu(x+\cT)=\pl A_\mu(x)\cP_\infty^{-1}$.
\begin{figure}[htb]
\vspace{5.7cm}
\includegraphics{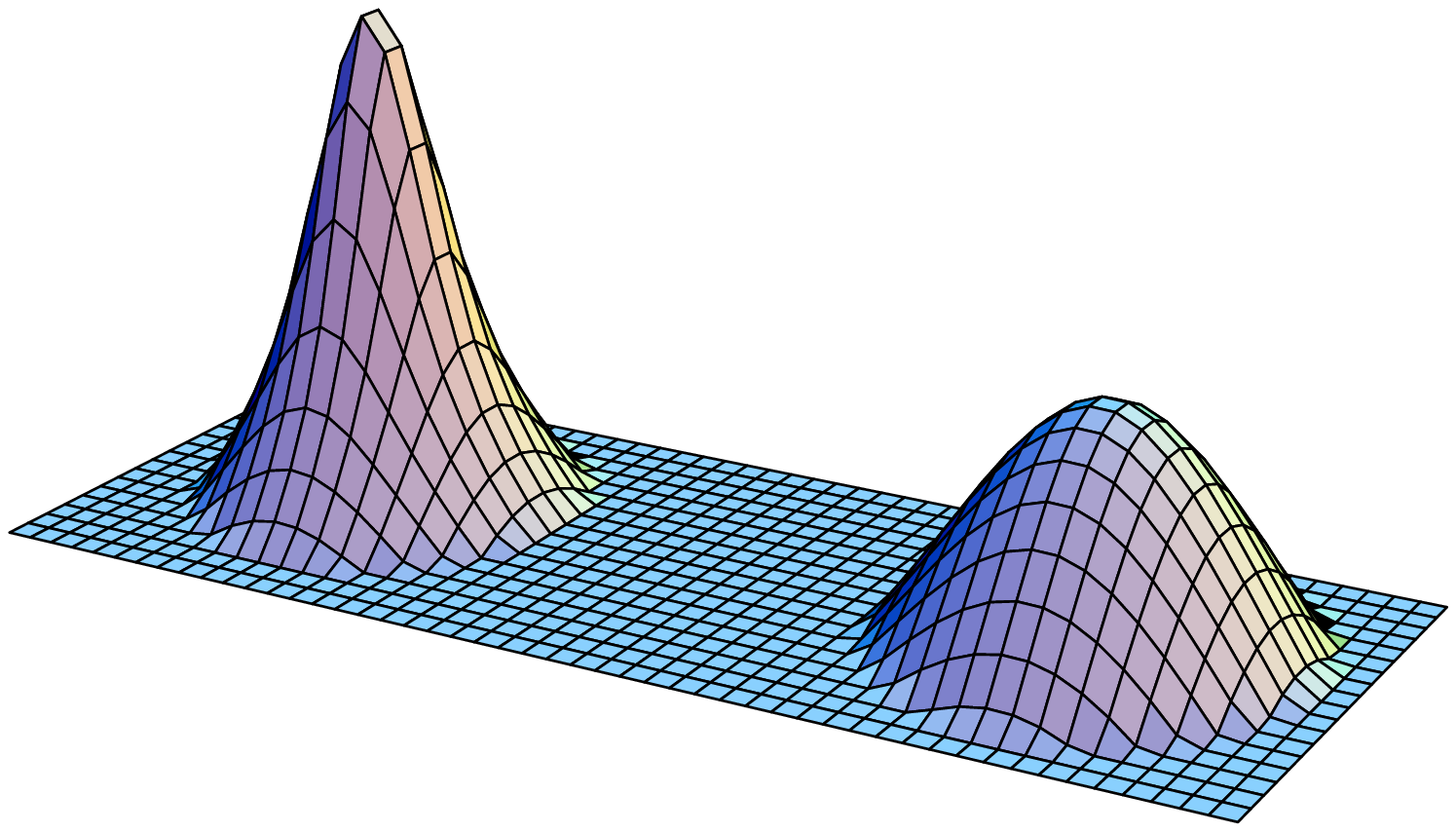}
\includegraphics{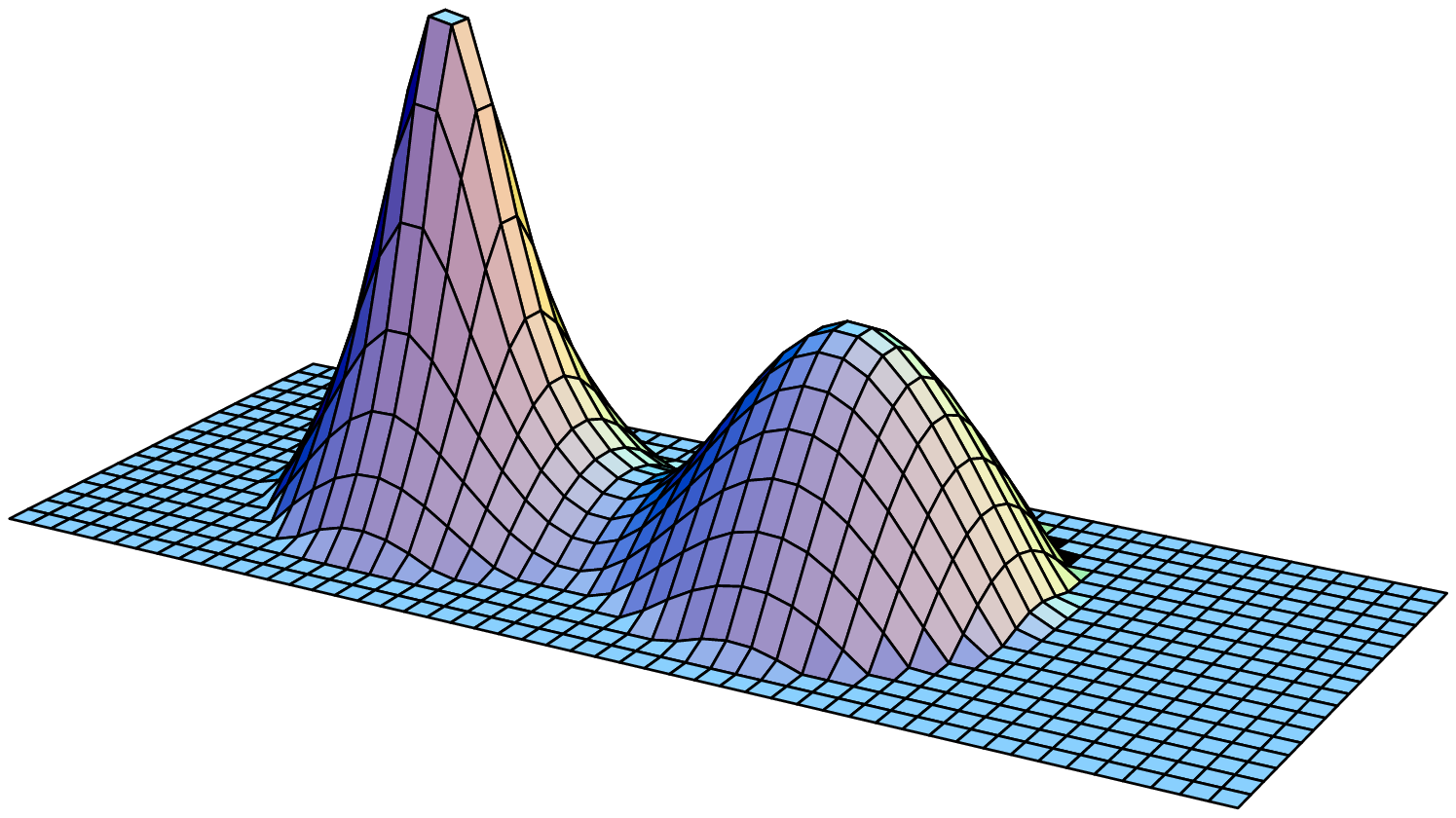}
\includegraphics{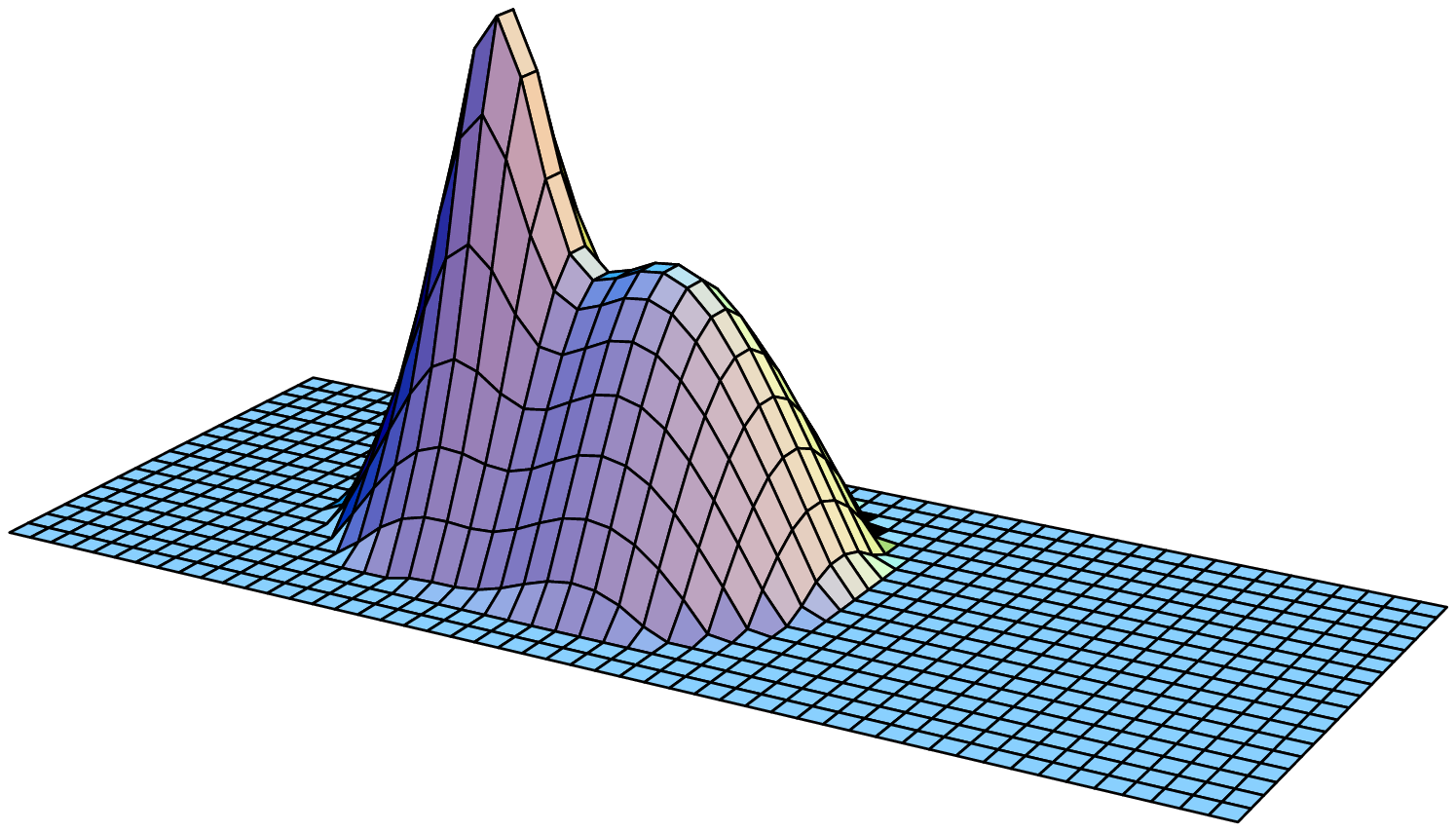}
\caption{Action densities for the $SU(2)$ caloron on equal logarithmic scales, 
cut off below $1/e^2$, for $t=0$, $\omega=0.125$, $\beta=1$ and
$\rho=1.6$ $1.2$ and $0.8$ (from top to bottom).}
\vspace{2.5cm}
\includegraphics{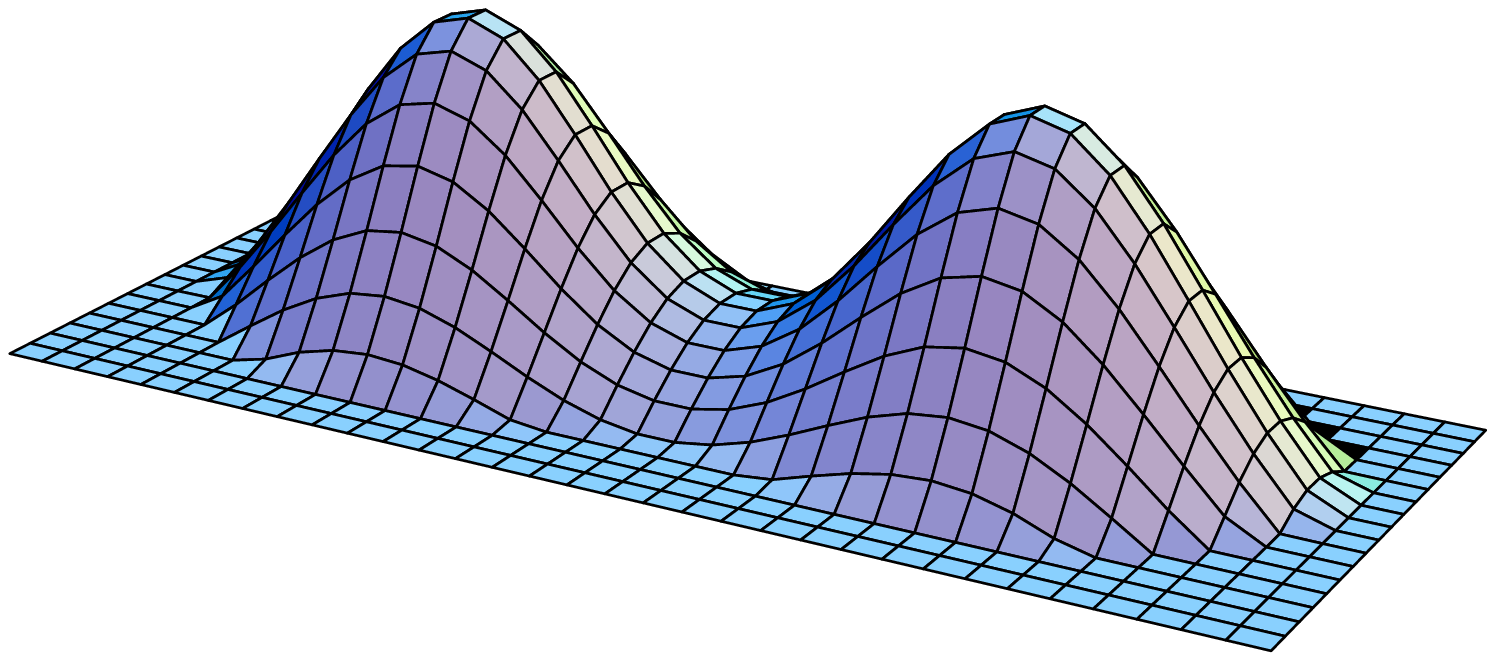}
\includegraphics{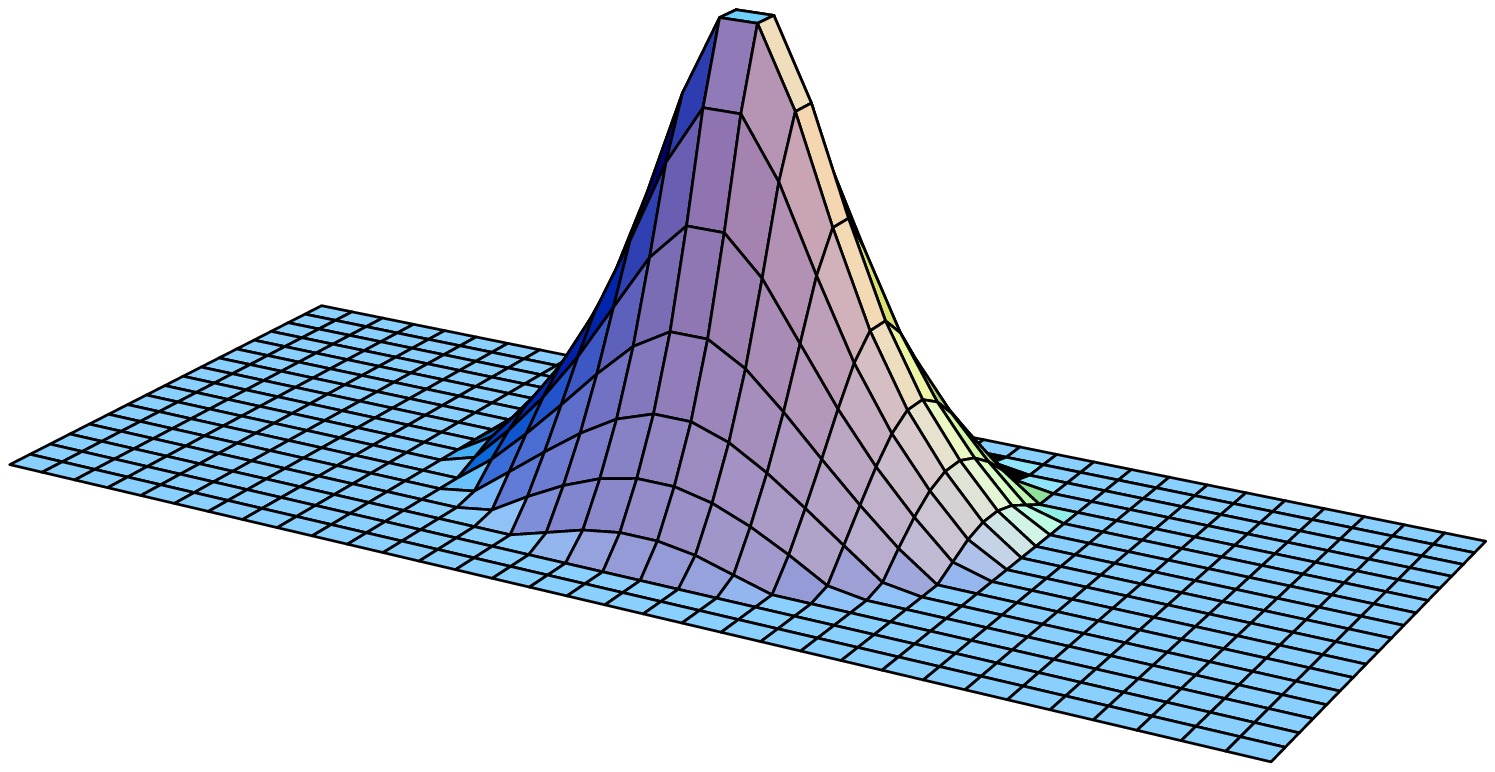}
\caption{As in fig.~2, now cut off below $1/e$, for $t\!=\!0$, 
$\rho\!=\!\beta\!=\!1$ with $\omega\!=\!\quarter$ (top) and $0$ (bottom).}
\end{figure}

For small $\rho$, equivalent to large $\cT$, the caloron approaches the 
ordinary single instanton solution, with no dependence on $\pl$. Finite size 
effects set in roughly when $\rho=\half\cT$. At this point, for $\nu_i\neq0$,
two lumps ($n$ for $SU(n)$) are formed, whose separation grows as $\pi\rho^2/
\cT$. When $\pl\!\!=\!(\!-\!)1$ for $SU(2)$, one of the lumps disappears, as 
$\nu_{1(2)}\!=\!0$, and the spherically symmetric Harrington-Shepard solution 
is retrieved. 

A non-trivial value of $\pl$ will modify the vacuum fluctuations and 
thereby leads to a non-zero vacuum energy density~\cite{GroPisYaf} as 
compared to $\cP\in Z_n$. A dilute, semi-classical instanton calculation 
is no longer considered a reliable starting point for QCD. Rather, it
is the monopole constituent nature from which we should draw important 
lessons for QCD~[5b].

\section{Monopoles from instantons}

At small $\cT$ the solution becomes static and the lumps are well separated 
and spherically symmetric. As they are \sd, they are necessarily BPS 
monopoles~\cite{BPS}. Also, when sending (at fixed $\beta$) one of the 
constituents to infinity, $|\vec y_m|\!\rightarrow\!\infty$, the solution 
becomes static and yields a simple way to obtain $SU(n)$ monopole 
solutions~[5c]. Explicitly we find (assuming $\nu_n\neq0$) in the 
limit $|\vec y_n|\!\rightarrow\!\infty$, which removes the $n$-th constituent, 
\beq
A_n\!\rightarrow\!2c_n\!\left(\!\!\!\bea{cc}1&\!\!\!1\\0&\!\!\!0\eea\!\!\!
\right),\quad A_{n\!-\!1}\!\rightarrow\!\frac{|\vec y_n|}{r_{n\!-\!1}}
\left(\!\!\!\bea{cc}s_{n\!-\!1}&\!\!\!c_{n\!-\!1}\\s_{n\!-\!1}&\!\!\!
c_{n\!-\!1}\eea\!\!\!\right),
\eeq
implying $\psi(x)\!\rightarrow\!2|\vec y_n|\exp(2\pi\nu_n|\vec y_n\!-\!\vec x|)
\tilde\psi(\vec x)$, with
\beq 
\tilde\psi(\vec x)=\half\tr\left\{\frac{1}{r_{n\!-\!1}}\left(\!\!\!\bea{cc}
s_{n\!-\!1}&\!\!\!c_{n\!-\!1}\\0&\!\!\!0\eea\!\!\!\right)\prod_{m=1}^{n\!-\!2}
\!A_m\right\}.
\eeq
As was emphasised in ref.~[5c], the energy density of the $SU(n)$ monopole is 
easily found from eq.~(3) (for a detailed description of some special cases 
see ref.~\cite{CLu})
\beq
{\cal E}(\vec x)=-\half\tr F_{\mu\nu}^{\,2}(\vec x)=-\half\Delta^2
\log\tilde\psi(\vec x).
\eeq

\section{Instantons from monopoles}

The new caloron solutions provide examples of gauge fields with 
topological charge built out of monopole fields, a construction going back to 
Taubes~\cite{Taubes}.  Non-trivial $\su2$ monopole fields are classified by 
the winding number of maps from $S^2$ to $SU(2)/U(1)\!\sim\!S^2$, where $\u1$ 
is the unbroken gauge group. Isospin orientations for a configuration made out
of monopoles with opposite charges behave as shown in fig.~4 (top), in a 
suitable gauge and sufficiently far from the core of both monopoles. 
\begin{figure}[b]
\vspace{3.3cm}
\includegraphics{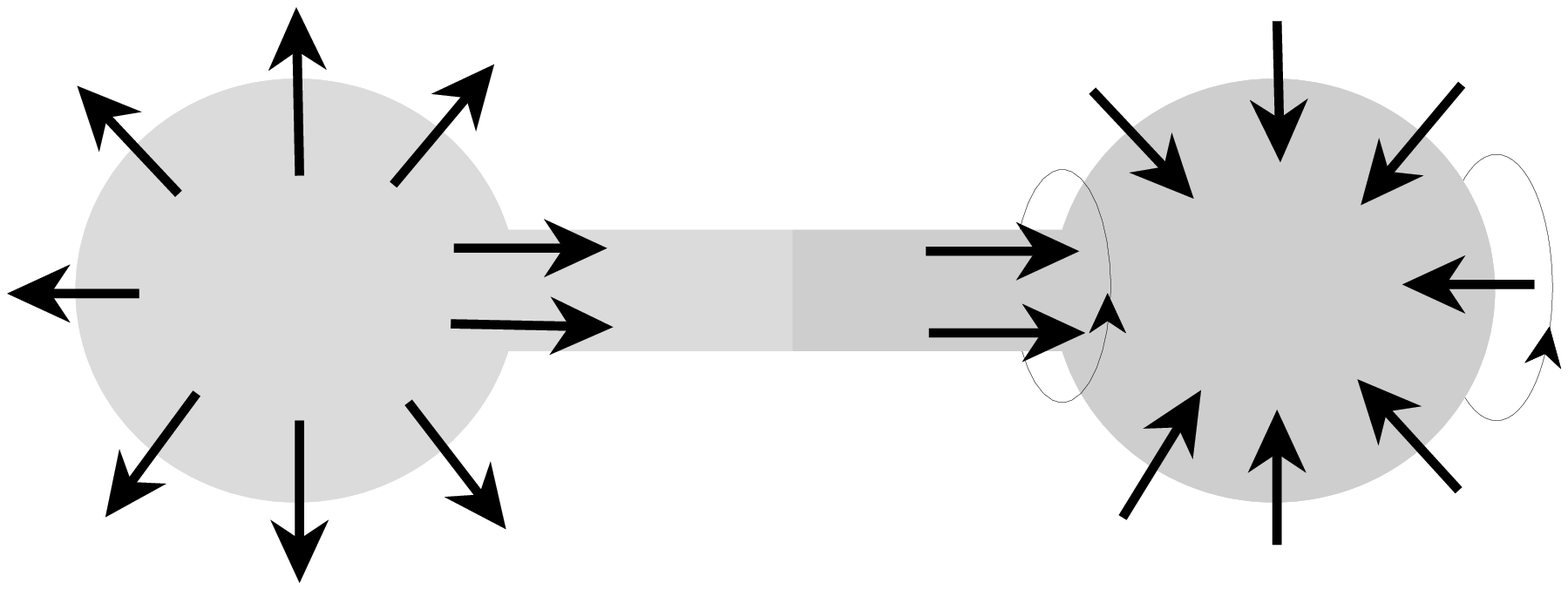}
\includegraphics{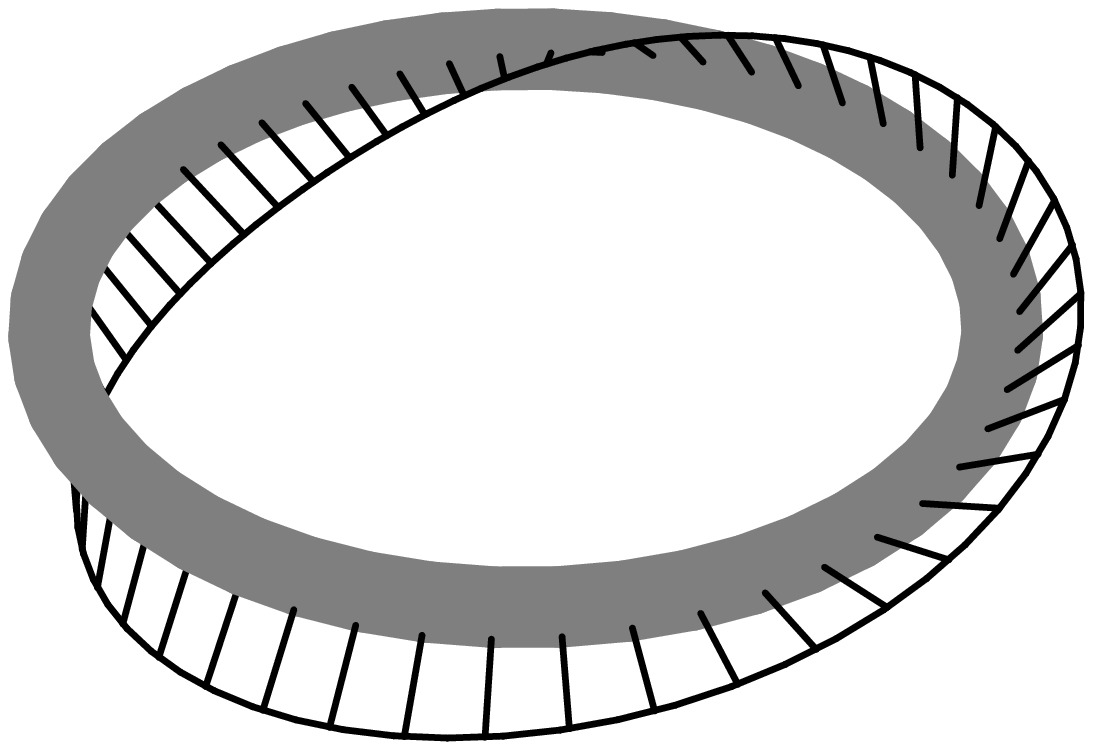}
\caption{Topological charge constructed from oppositely charged
monopoles by rotating one of them. For a closed monopole line, the 
embedding of the unbroken subgroup makes a full rotation.}
\end{figure}
Taubes constructed
topologically non-trivial configurations by creating a monopole anti-monopole 
pair, bringing them far apart, rotating one of them over a full rotation and 
finally bringing them together to annihilate (cmp. fig.~5). We can describe 
this as a closed monopole line (or loop) with the orientation of the core 
defined by $SU(2)/U(1)\!\sim\!S^2$, ``twisting'' along the loop, thus 
describing a Hopf fibration~[5b] (see fig.~4 (bottom)). The only topological 
invariant available to characterise the homotopy type of this Hopf fibration 
is the Pontryagin index. It prevents full annihilation of the ``twisted''
monopole loop.

For large $\rho$, eq.~(4) gives up to exponential corrections, i.e.
outside the cores of the constituents, 
\beq
A_\mu=\frac{i}{2}\tau_3\bar\eta^3_{\mu\nu}\partial_\nu
\log\phi_0,\quad\phi_0\equiv\frac{r_1+r_2+\pi\rho^2}{r_1+r_2-\pi\rho^2}. 
\eeq
This describes two Abelian Dirac monopoles and one easily verifies $\log\phi_0$
is harmonic, as required by self-duality. Furthermore $\phi_0^{-1}$ vanishes 
on the line connecting the two monopole centers, giving rise to return flux, 
absent in the full theory. The relative phase $e^{-2\pi it}$ in the expression 
for $\chi$ given before, describes the full rotation of the core of a 
constituent monopole, required so as to give rise to non-trivial topology.

A conjectured QCD application, in the form of a hybrid monopole-instanton 
liquid, was discussed in ref.~[5b]. Abelian projection applied to our 
solutions was also discussed at this conference~\cite{BroNeg}.
\begin{figure}[htb]
\vspace{2.2cm}
\includegraphics{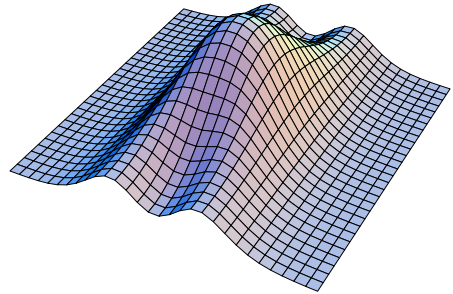}
\caption{Action density in the $z$-$t$--plane for $x\!=\!y\!=\!0$, $\omega\!=\!
\quarter$, $\rho\!=\!\half$ and $\beta\!=\!1$ on a linear scale. One can 
trace the constituent monopoles in the low field regions, ``annihilating'' to
give an instanton.}
\end{figure}
${}$
\vskip-5mm
${}$

\end{document}